# Ultra Compact low cost two mode squeezed light source

Shahar Monsa, Shmuel Sternklar, Eliran Talker

Department of Electrical Engineering, Ariel University, Ariel 40700, Israel

*Corresponding author: elirant@ariel.ac.il

## Abstract

Quantum-correlated states of light, such as squeezed states, constitute a fundamental resource for quantum technologies, enabling enhanced performance in quantum metrology, quantum information processing, and quantum communications. The practical deployment of such technologies requires squeezed-light sources that are compact, efficient, low-cost, and robust. Here we report a compact narrowband source of two-mode squeezed light at 795 nm based on four-wave mixing in hot $^{85}$Rb atomic vapor. The source is implemented in a small, modular architecture featuring a single fiber-coupled input, an electro-optic phase modulator combined with a single Fabry–Perot etalon for probe generation, and two free-space output modes corresponding to the signal and conjugate fields. Optimized for low pump power, the system achieves up to −8 dB of intensity-difference squeezing at an analysis frequency of 0.8 MHz with a pump power of only 300 mW. The intrinsic narrowband character of the generated quantum states makes this source particularly well suited for atomic-based quantum sensing and quantum networking, including interfaces with atomic quantum memories. Our results establish a versatile and portable platform for low-SWaP squeezed-light generation, paving the way toward deployable quantum-enhanced technologies.

## Introduction

Squeezed states of light have become an indispensable resource for quantum sensing, quantum communication, and quantum information processing [1–8]. By reducing quantum noise below the standard quantum limit relative to classical coherent states, squeezed light enables enhanced measurement sensitivity and precision [7]. A prominent example is the use of squeezed light to improve the sensitivity of large-scale laser interferometers such as the Laser Interferometer Gravitational-Wave Observatory (LIGO), leading to major advances in gravitational-wave detection [9–12]. Beyond metrology, squeezed states play a central role in continuous-variable quantum computing [13,14], high-fidelity quantum teleportation [15], and secure quantum communication protocols [16] .For squeezed light to transition from laboratory demonstrations to real-world applications, it is essential to develop sources that are robust, compact, and cost-effective, with low size, weight, and power consumption (SWaP). Deployable quantum technologies demand optical architectures that minimize complexity, optical loss, and excess noise while maintaining high squeezing performance. Conventional approaches based on Mach–Zehnder electro-optic modulators (MZ-EOMs) typically require two LiNbO$_3$ crystals, an active bias controller, and multiple Fabry–Perot (FP) etalons often three or more to suppress unwanted modulation sidebands [17,18]. These requirements significantly increase system cost, footprint, and power consumption. Similarly, double-pass acousto-optic modulator (AOM) schemes suffer from low diffraction efficiency (typically <10%) and require additional optical components that introduce loss and excess intensity noise, thereby degrading the quality of the generated probe beam. Significant progress has been made in the development of modular squeezed-light sources using optical parametric oscillators [19], parametric down-conversion in nonlinear crystal waveguides (both free-space [20] and fiber-coupled [21]), and four-wave mixing (FWM in optical fibers [22]). However, comparatively little effort has been devoted to the realization of low-cost, low-SWaP, narrowband squeezed-light sources. Such sources are of particular importance for atomic-based quantum technologies including quantum memories [23,24], quantum sensors [25–27], and atomic clocks [28,29] and optical clocks [30,31] where efficient light–matter interaction requires the optical bandwidth of the quantum state of light to be well matched to the narrow linewidths of atomic transitions. As the push toward deployable and field-ready quantum systems continues, the development of compact, narrowband, low-SWaP sources of nonclassical light becomes increasingly critical.

In this work, we present a compact and low-cost narrowband squeezed-light source that addresses these challenges. In contrast to previous implementations [22,32], which either neglect the impact of probe-beam generation or rely on double-pass AOM or MZ-EOM combined with multiple FP etalons, we employ a single electro-optic phase modulator (EOPM) in conjunction with only one FP etalon to generate the frequency-shifted probe beam. This simplified architecture significantly reduces optical loss and excess noise while eliminating the need for interferometric stabilization and bias control. We experimentally demonstrate that this approach yields a lower initial probe-beam noise compared to both double-pass AOM and MZ-EOM schemes and enables the generation of near −8 dB of two-mode squeezing via four-wave mixing in hot rubidium vapor.

**Theoretical background**

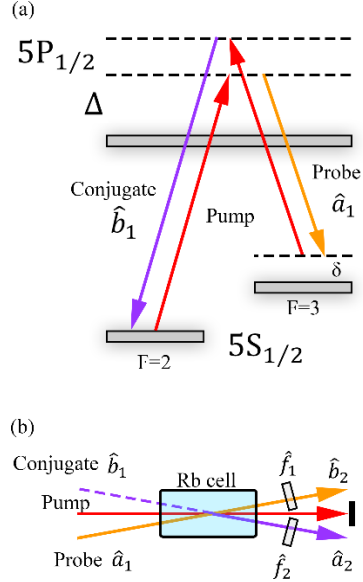

**FIG. 1.** Double-Λ scheme. (a) Energy-level diagram of the $^{85}$Rb D1 line used in the double-Λ configuration. $\Delta$ denotes the one-photon detuning, and $\delta$ denotes the two-photon detuning. (b) Schematic of the double-Λ four-wave-mixing interaction. The pump, probe, and conjugate optical fields are shown in red, yellow, and violet, respectively, for clarity.

A double-Λ four-wave mixing (FWM) process in hot rubidium vapor is employed, as schematically illustrated in Fig. 1. The corresponding configuration realizes a phase-insensitive amplifier (PIA) based on the FWM interaction, as shown in Fig. 1. The PIA is driven by a strong coherent pump field, while the probe and conjugate input ports are seeded by a weak coherent field $\hat{a}_1$ and a vacuum field $\hat{b}_1$, respectively. Through the FWM process, two pump photons are annihilated to generate a pair of correlated photons at the probe (signal) and conjugate frequencies. As a result, the probe beam is amplified, and a new conjugate beam is generated. When the optical fields propagate through the atomic vapor, they experience not only parametric gain but also internal losses, which are dominated by atomic absorption. To model these losses, two fictitious beam splitters with identical transmissivity $\eta_1$ are introduced after the vapor cell. The vacuum modes associated with these losses are denoted by

$\hat{f}_1$ and $\hat{f}_2$. Under this model, the output field operators $(\hat{a}_2, \hat{b}_2)$ of the PIA can be written as

(1)
$$\begin{pmatrix} \hat{a}_2 \\ \hat{b}_2^\dagger \end{pmatrix} = \sqrt{\eta_1} \begin{pmatrix} \sqrt{G} & -\sqrt{G-1} \\ -\sqrt{G-1} & \sqrt{G} \end{pmatrix} \begin{pmatrix} \hat{a}_1 \\ \hat{b}_1^\dagger \end{pmatrix} + \sqrt{1-\eta_1} \begin{pmatrix} \hat{f}_1 \\ \hat{f}_2^\dagger \end{pmatrix}$$

where G is the intensity gain of the FWM interaction. The degree of intensity-difference squeezing between the output probe and conjugate beams, normalized to the standard quantum limit (SQL), is given by

(2)
$$DS = \frac{Var(\hat{N}_{a_2} - \hat{N}_{b_2})_{PIA}}{Var(\hat{N}_{a_2} - \hat{N}_{b_2})_{SQL}} = \eta_1 \frac{1}{2G-1} + 1 - \eta_1$$

where $\hat{N}$ denotes the photon-number operator and the subscripts refer to the corresponding optical modes. Equation (2) shows that, ideally, the achievable squeezing is independent of the initial noise of the probe beam. However, in practice, the probe beam must be frequency-detuned by 3.04 GHz from the pump beam to satisfy the two-photon resonance condition in $^{85}$Rb. This frequency shift typically requires additional optical components that introduce excess intensity noise.

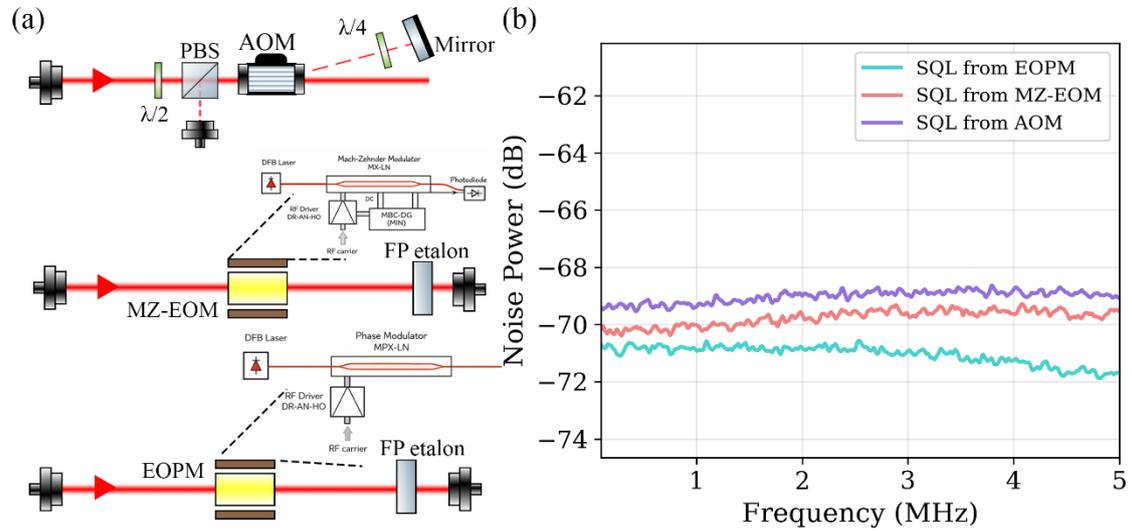

FIG 2. (a) three probe beam generators method including double pass AOM, MZ-EOM with a single FP etalon, and EOPM follows by a single FP etalon (b) measured SQL noise of the three probe beam generators

Consequently, even after the squeezing process, the effective two-mode squeezing can be significantly degraded, limiting the practical performance of FWM-based quantum light sources in rubidium vapor. Several approaches have been explored to generate the frequency-detuned probe beam. The first method employs two independent lasers phase-locked using a phase-locked loop (PLL), as demonstrated in Ref. [18], where up to −6 dB of squeezing was reported. The reduced squeezing in this approach is attributed to excess intensity noise introduced by the PLL, which directly modulates the diode-laser current. A second method uses a double-pass acousto-optic modulator (AOM), as illustrated in Fig. 2(a), where a mirror and a polarizing beam splitter (PBS) are used to couple the frequency-shifted beam into an optical fiber [33–36]. A third approach relies on a MZ-EOM followed by three Fabry–Perot (FP) etalons before fiber coupling [17,18,22]. In contrast, our method employs a EOPM in combination with only one FP etalon. As shown in Fig. 2(b), this configuration yields the lowest initial probe-beam noise among all the tested approaches. This improvement can be attributed to the reduced number of optical components and the absence of interferometric structures. Compared to the MZ-EOM scheme, which requires two LiNbO$_3$ crystals and a Mach–Zehnder interferometer both of which introduce additional loss and excess noise our EOPM-based approach minimizes optical loss and preserves the quantum correlations more effectively.

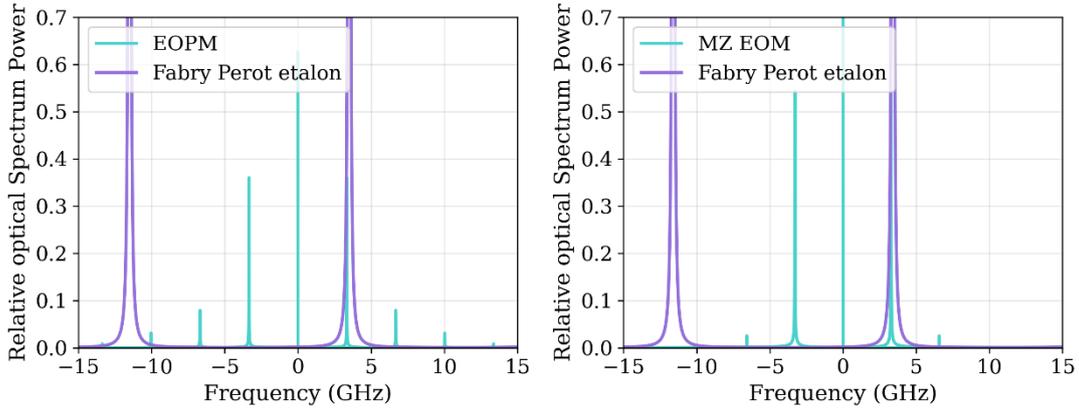

**FIG. 3.** Optical power spectra (turquoise) generated using an electro-optic phase modulator (EOPM) and a Mach–Zehnder electro-optic modulator (MZ-EOM), shown in comparison with the transmission peaks of a Fabry–Perot (FP) etalon (violet).

Next, we analyze the efficiency of generating the frequency-shifted probe beam after transmission through a single Fabry–Perot (FP) etalon. The electric field at the output of the EOPM can be written as

$$E(t) = \frac{1}{2}E_0 e^{i\omega_0 t}[J_0(\alpha) + 2J_2(\alpha)\sin(2\Omega t) + 2J_3(\alpha)\sin(3\Omega t) + \cdots] \quad (3)$$

where $E_0$ is the input field amplitude, $\omega_0$ is the optical carrier frequency, $\Omega$ is the RF modulation frequency, $\alpha$ is the modulation index, and $J_n(\alpha)$ denotes the $n$-th order Bessel function of the first kind. For comparison, the output electric field obtained using a MZ-EOM can be expressed as

$$E(t) = \frac{1}{2} E_0 e^{i\omega_0 t}[4J_1(\alpha)\sin(\Omega t) + 4J_3(\alpha)\sin(3\Omega t) + \cdots] \quad (3)$$

Fig. 3 shows the corresponding optical power spectra, $P = EE^*$, for both the EOPM and the MZ-EOM configurations. The transmission peaks of a Fabry–Perot etalon with a free spectral range (FSR) of 15 GHz are indicated by the purple line. Owing to the large spectral separation between adjacent modulation sidebands, the FP etalon efficiently suppresses all unwanted frequency components, while selectively transmitting the desired sideband. As a result, we demonstrate that a single electro-optic phase modulator combined with a single FP etalon is sufficient to generate a clean probe beam frequency-shifted by 3.04 GHz. This approach avoids the complexity and excess loss associated with interferometric modulators, while maintaining low added noise and high spectral purity, making it particularly well suited for low-noise FWM-based quantum light generation in rubidium vapor.

**Experimental setup and results**

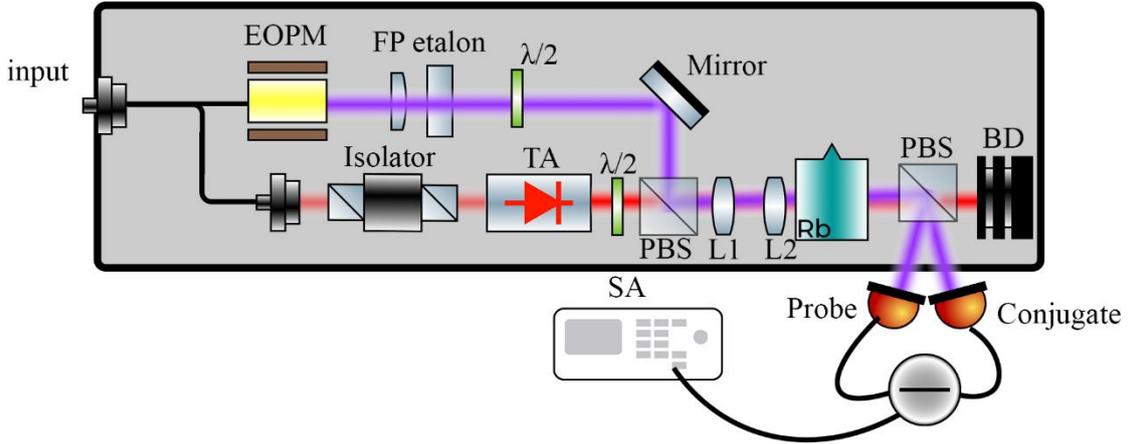

**FIG. 4**. Experimental setup for the ultra-compact double-Λ configuration based on the $^{85}$Rb D1 line. TA, tapered amplifier; M, mirror; L, lens; PBS, polarizing beam splitter; EOPM, electro-optic phase modulator; λ/2, half-wave plate; SA, spectrum analyzer.

We generate narrowband two-mode squeezed light at 795 nm using a four-wave mixing (FWM) process in a double-Λ configuration in hot $^{85}$Rb vapor. In this nonlinear interaction, two pump photons at frequency $\omega_p$ and one seed photon at the probe frequency $\omega_{pr}$ interact with the atomic medium, resulting in the simultaneous emission of correlated photon pairs: one at the probe frequency and one at the conjugate frequency $\omega_c$, as required by energy conservation. Consequently, the probe field is amplified while a conjugate field is generated. The strong pump beam is derived from an external-cavity diode laser (ECDL) operating at 795 nm and amplified using a tapered amplifier. A small fraction of the pump light is frequency down-shifted by 3.04 GHz—corresponding to the hyperfine splitting between the $|5S_{1/2}, F = 2\rangle$ and $|5S_{1/2}, F = 3\rangle$ ground states—using a waveguide EOPM. A single FP etalon with a free spectral range of 15 GHz and a finesse exceeding 30 is used to spectrally filter the modulation sidebands, producing a clean probe beam. As a result, the pump and probe frequencies are tuned close to the two-photon Raman resonance between the ground-state hyperfine levels. The correlated generation of probe and conjugate photons leads to strong temporal quantum correlations between their intensity fluctuations, enabling noise levels below the shot-noise limit (SNL). These correlations are characterized by measuring the optical powers of the probe and conjugate beams with a balanced photodetector, subtracting the corresponding photocurrents, and analyzing the resulting intensity-difference noise using a spectrum analyzer. The FWM interaction requires orthogonally polarized pump and probe beams that spatially overlap inside the atomic vapor at a small angle. To ensure stable spatial modes and well-defined orthogonal polarizations, the pump and probe beams are coupled into two independent single-mode polarization-maintaining fibers. After exiting the fibers, the beams intersect at an angle of approximately 0.5° at the center of a 12.5-mm-long isotopically pure $^{85}$Rb vapor cell with anti-reflection-coated windows, heated to approximately 120°C. Following the FWM interaction, a polarizing beam splitter suppresses residual pump light, and the probe and conjugate beams are directed to a balanced detector for intensity-difference noise measurements. The output of this photodetector is fed into a radio frequency spectrum analyzer with a resolution bandwidth (RBW) of 30 kHz and a video bandwidth (VBW) of 300 Hz. In addition, we introduce a delay line into the conjugate beam path to compensate for the differential slow-light delay discussed in Ref. [33,34].

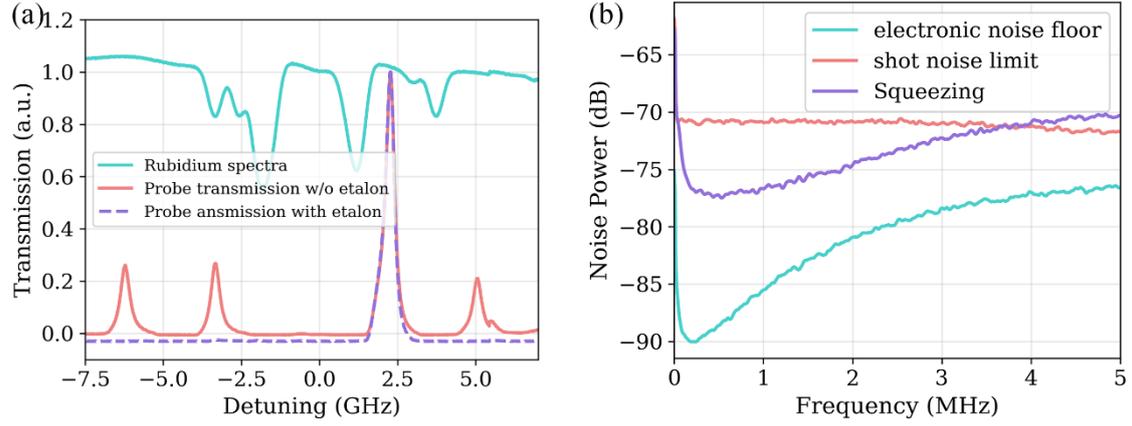

**FIG. 5.** (a) Probe transmission spectrum after the electro-optic phase modulator (EOPM) with a single Fabry–Perot (FP) etalon (violet) and without the FP etalon. The turquoise trace shows the transmission spectrum of the natural $^{85}$Rb D1 line. (b) Measured intensity-difference squeezing on the $^{85}$Rb D1 line, compared to the electronic noise floor (turquoise) and the shot-noise limit (SQL, red).

Figure 5(a) shows the measured probe transmission spectrum with and without the FP etalon. When the EOPM is driven at 3.040 GHz with a modulation amplitude of 2.05 dB and an input probe power of 18 mW, unwanted modulation sidebands are strongly suppressed by the FP etalon, leaving a single clean probe frequency. The etalon temperature is actively stabilized using a PID controller to maintain optimal filtering. By tuning the one-photon detuning to 0.9 GHz and the two-photon detuning to −8 MHz, optimal squeezing conditions are achieved. Figure 5(b) shows the measured noise power spectral density of the intensity-difference signal. A maximum noise reduction of approximately −8 dB below the SNL is observed in the low-frequency range from 40 to 500 kHz, with a squeezing bandwidth extending to about 3 MHz. These results are obtained for a pump power of 300 mW and a seed probe power of 10 μW. Under these conditions, the probe beam is amplified to 150 μW (gain $G \approx$ 15), while the conjugate beam reaches a power of 120 μW, resulting in a total detected optical power of 270 μW. To quantitatively evaluate the degree of squeezing and assess the contribution of system noise, we vary the seed probe power and measure the intensity-difference noise as a function of total optical power.

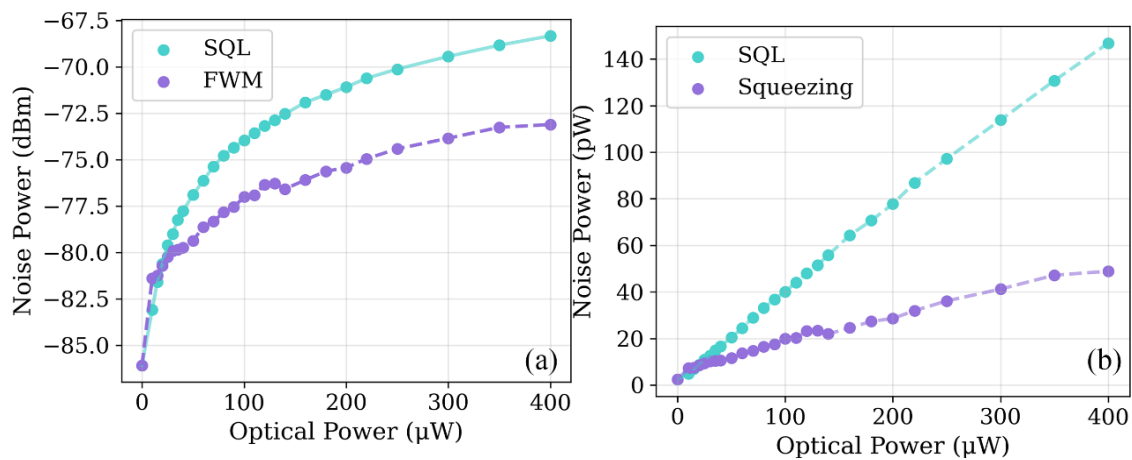

**FIG 6**. (a) Relative intensity noise measured at ∼ 1MHz as a function of noise power. (b) Relative intensity noise at ∼ 1MHz for the shot-noise limit (SQL) and the four-wave-mixing (4WM) configuration, plotted as a function of the total optical power incident on the photodetector.

Figure 6(a) shows the measured noise power (in dBm) for both the SQL and the FWM-generated twin beams. While the SQL increases with optical power, the intensity-difference noise of the twin beams exhibits a significantly reduced slope, indicating the presence of strong quantum correlations. Figure 6(b) presents the same data in linear units, clearly highlighting the different scaling behaviors. The ratio between the slopes of the squeezed-state noise and the SQL is extracted to be 0.135, corresponding to approximately −8 dB of intensity-difference squeezing. Comparing this value with the directly measured squeezing level of −8 dB in Fig. 2(b), we estimate that system noise contributes approximately 1 dB of degradation. The dominant sources of system noise are attributed to residual pump scattering and the electronic noise of the detection system. Under the present operating conditions, optical and technical losses are close to the limit that determines the maximum achievable squeezing in our setup.

**Conclusion**

In conclusion, we have demonstrated a compact, low-cost, and low-SWaP narrowband squeezed-light source based on four-wave mixing in hot rubidium vapor, enabled by a fundamentally simplified probe-beam generation architecture. By replacing interferometric and acousto-optic frequency-shifting schemes with a single electro-optic phase modulator combined with a single Fabry–Perot etalon, we drastically reduce optical complexity, loss, and excess technical noise key factors that have historically limited the practicality of atomic vapor–based squeezed-light sources. We show that this minimal architecture produces a probe beam with significantly lower initial intensity noise than state-of-the-art implementations relying on double-pass AOMs or Mach–Zehnder electro-optic modulators with multiple filtering etalons. As a direct consequence, we achieve near -8 dB of two-mode intensity-difference squeezing, approaching the fundamental limits set by internal atomic loss. Importantly, our results demonstrate that the achievable squeezing is no longer constrained by probe-generation noise, but rather by intrinsic properties of the nonlinear medium marking a decisive step toward truly optimized atomic FWM systems. Beyond performance metrics, the significance of this work lies in its practical impact. The elimination of interferometric bias control, multiple $LiNbO_3$ crystals, and high-power RF-driven AOM stages results in a system that is not only simpler and more robust, but also substantially more open to integration, miniaturization, and field deployment. This directly addresses a long-standing gap between laboratory-scale squeezed-light experiments and real-world quantum technologies. The narrowband nature of the generated quantum states, inherently matched to atomic transitions, makes this source particularly well suited for atomic quantum memories, quantum-enhanced sensors, precision spectroscopy, and next-generation atomic clocks. More broadly, our approach establishes a clear design principle: high-quality quantum light does not require complex or resource-intensive modulation schemes, but can instead be realized through carefully engineered, low-noise minimal architectures. We therefore anticipate that this work will serve as a blueprint for the development of deployable atomic-based quantum systems, enabling the transition of squeezed-light–enhanced technologies from controlled laboratory environments to practical applications in navigation, timing, sensing, and communication.